\documentclass[conference]{IEEEtran}

\usepackage{cite} 
\usepackage[pdftex]{graphicx}
\graphicspath{{./images/}{./figures/}}
\DeclareGraphicsExtensions{.pdf,.jpeg,.png}

\usepackage{amsmath}
\usepackage{algorithmic}
\usepackage{array}
\usepackage{url} 
\usepackage{hyperref}
\usepackage{url}        

\usepackage{amsmath}    

\usepackage{textgreek}	
\usepackage{listings}
\usepackage{csvsimple}
\usepackage{longtable}
\usepackage{cuted}
\usepackage{caption}
\usepackage{booktabs}
\usepackage{tabularx}

\title{Carbon Dioxide Emission Minimized Virtual Machine (VM) Placement in Cloud-Fog Network Architecture}

\author{
  Tarek Bessalah \\
}

\begin{document}
\maketitle

\begin{abstract}
Cloud computing has provided economies of scale, savings, and efficiency for both individual consumers and enterprises. Its key advantage is its ability to handle increasing amounts of data and provide functionality that gives users the ability to scale their computing resources, including processing, data storage, and networking capabilities. Virtual Machines (VM), enabled via virtualization technology, allow cloud service providers to deliver their services to users. This, however, results in increasing carbon dioxide emissions from increased energy use.

This paper introduces a Mixed-Integer Linear Programming (MILP) model that investigates the VM placement, focusing on the British Telecom (BT) network topology, in a cloud-fog network architecture when renewable energy sources are introduced in the fog layer located near traffic-producing sources. VMs can be placed on nodes hosted on the core, metro, and access (fog) layers. We first investigate the effect of varying traffic on IP over WDM power consumption in the backbone network and the number of optical carrier signals to serve the traffic over a period of time. We later extend the model to consider the CO2-minimized optimal virtual machine placement given the
sporadic traffic quantity, and the consideration of solar renewable energy
sources placed in data centers located in the access (fog) layer throughout the day, imposed on the VM and the minimum workload requirement of the VM to maintain a service-level agreement (SLA).

This paper suggests a direct proportionality between power consumption and the imposed traffic. When integrating solar power into data centers within the British Telecom (BT) network topology, there was a noticeable reduction in power consumption, amounting to up to 16 percent in nodes that received solar energy. This paper demonstrates that a diminishing VM workload led to decreased VM replication in the metro and access layers for constant profiles. The linear profile exhibited the inverse behavior.
\end{abstract}

\section{Introduction}
The primary selling point of cloud computing is its ability to accommodate exponential growth in voluminous data and provide entities
to host functions by providing access, seemingly boundless, to a continuum of computing resources such as processing, data storage, and
networking functions. The virtualization process, utilized by cloud service providers, produces VM services that allow for such functionality \cite{Odun-Ayo2017}.
With the ever-increasing demand for cloud computing, strain has been imposed on cloud data centers and the energy utilized from the equipment
needed to maintain the computing equipment is increasing. CO2 emissions are directly correlated to the electricity demand, which is expected to
increase by 15-30\% in 2025 \cite{IEA2022}.

As an attempt to alleviate the exponentially increasing
energy expenditure and the strain imposed on data centers
brought by traffic, fog computing is a paradigm that makes
available computing components, that are able to host virtual
machines, placed closer to traffic producing sources. The fog
layer brings forth functionality such as filtering, in which
voluminous traffic can be filtered in the fog layer without
having redundant data be sent to the cloud layer, thus
providing transport network energy and bandwidth savings
\cite{Qu2020}. Fog computing, in this paper, indicates the data centers
located nearby traffic producing sources in the access network.

\section{Related Work and Background}
\begin{figure}
    \centering
    \includegraphics[width=0.7\linewidth]{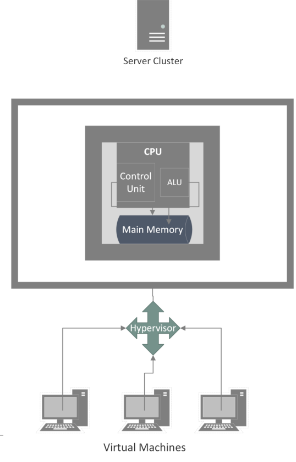}
    \caption{Multiple VM placement on a single server cluster node}
    \label{fig:Multiple-VMs}
\end{figure}
Numerous studies have been conducted to further optimize
energy usage in the fog layer. In \cite{AliKumar2022}, the authors proposed a
Green-Demand Aware Fog Computing (GDAFC) solution
that uses a prediction technique to map fog server nodes to
different work profiles: working nodes, standby nodes, and
idle nodes. A prediction model which predicts future volume
of requests based on previous trends allocates working nodes
for the precedented in, allowing nodes that are not utilized to
be placed in idle mode. This resulted in up to 65\% savings in
energy in the fog layer. In \cite{Cui2019}, the authors investigate a
Lyapunov optimization technique, that derives a greedy
algorithm and a joint algorithm, which maps requests, in a
network queue, to the amount of energy required to fulfill the
requests and investigates whether to dispatch requests one by
one or all at once. In \cite{Musaddiq2020}, the authors investigated the energy
efficiency of three different packet routing based objective
functions, Objective Function Zero (OF0), Advanced
Objective Function Zero (AF0), and Minimum Rank with
Hysteresis Objective Function (MRHOF) for Routing
Protocol for Low Power and Lossy Networks (RPL) in the fog
layer. RPL is a routing protocol that assigns nodes a rank in
respect to the root node. OF0 and MRHOF formulate a route
based on hop count and expected transmission count
respectively. The authors proposed AF0, which routes based
on congestion, is optimal for QoS and
energy efficiency. It is worth noting that fog computing was
not introduced solely for its potential to save power, but to
provide users with increased quality of service and low latency
by offloading services on fog nodes whenever appropriate.

Cloud computing provides services that are in the form of
Infrastructure-as-a-Service (IaaS) \cite{RedHat2022}, allowing users access to
processors, routers, and data storage through means of
virtualization techniques that allow for data replication, log
access, and machinery health. Platform-as-a-Service (PaaS)
\cite{RedHat2022} provides users with an operating system, development
tools, and business analytics, that can be accessed through
manufacturer-specific API calls. Software-as-a-Service
(SaaS) \cite{RedHat2022} are applications that can be accessed remotely via
the internet, otherwise known as web-based software.
Virtualization is a technique that provides customers with
the mentioned services, by virtualizing underlying computing
resources, such as CPU, data storage, and networking tools, of
physical hardware to the extent the customer specifies. For example, customers can control the number
of cores and the size of memory. Figure
1 illustrates how multiple virtual machines can be placed on a
single server. 

Virtualization provides
interfaces to a variety of hardware-specific applications
enabling them to seemingly run on different platforms. This
especially comes to use in networking, where clusters
containing various platforms and functionality can be accessed
simultaneously by customers through means of a virtual
machine. Virtualization allows for entire core network service
functionalities to be placed in edge servers, within the
proximity of a subscriber, through means of a virtual machine copy, this, however, comes at the expense of increased computing power and, therefore increased emissions. In contrast, placing a virtual machine in a core server node comes at the expense of transport network processing costs.

\begin{figure*}
    \centering
    \includegraphics[width=1.05\textwidth]{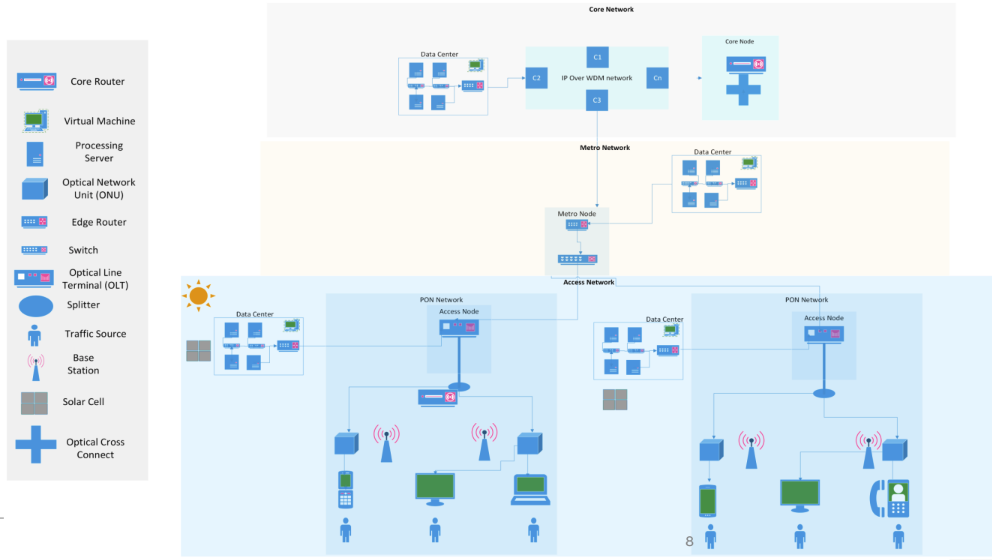}
    \caption{Cloud to fog architecture with solar power consideration.}
    \label{fig:solar-power-topology}
\end{figure*}

VMs provide IaaS, PaaS, and SaaS cloud functionalities via
VM categories such as Hardware Virtual Machine (HVM),
allowing a machine to be virtualized with no operating system,
and Container Virtual Machine, allowing multiple secure
containers to run different applications on a single kernel, and
facilitates user access to containers without interaction with
other users. In this paper, Container Virtual Machines
will be in consideration as it is the most efficient amongst other
virtualization categories \cite{Zhang2018}.

One might wonder, to what extent running a virtual machine
on a host machine would affect power consumption compared
to running it on native hardware. The difference in power
consumption is minuscule and, therefore negligent \cite{Stoess2007}. This is because
many of the virtualized services, such as the operating system
and functionality, can be directly processed by the underlying
host. Several papers have discussed different approaches to
ensure virtualization and allocation are performed to ensure
minimal power expenditure in mind. In \cite{Khani2022}, the authors propose an algorithm, that reduces the number of nodes, based
on best-fit, in which VMs with a certain threshold of high
traffic are placed on the same host machine to reduce traffic.
The algorithm traverses through a list of available VMs, for
each VM it searches for available computing nodes that can
allocate it, if none is available, then a new computing node is
allocated. 

An under-utilized server can consume up to 70\% of
run-time power \cite{MeisnerGoldWenisch2009}. In \cite{Gupta2022}, the authors investigate the dynamic reallocation of VMs according to CPU usage. This in turn becomes a bin-packing problem, where VMs are always evaluated on the basis of
whether they should be placed in other nodes and their current
node needs to be switched off. The power consumption of
the processing unit of the underlying server is directly
proportional to the workload imposed on the VMs \cite{Ismail2018}, with the CPU
component utilizing a greater portion of power compared to
the storage and networking functions of the server. Therefore, this
paper will consider two work profiles, constant and linear
CPU workload profiles.

 As data center components, such as routers, switches, and
servers, etc. become more prone to failure with increased traffic \cite{Machida2010}, VM replication schemes have been widely adopted to increase overall redundancy and distribute
workload tasks. Related research in this area focus primarily
on the server consumption with hosting VMs, however, this
work considers the power consumption of additional
networking components such as routers and switches.
It must be noted that this work only considers the
CPU capacity, and not the limitations with the storage capacity of
the server.

The consideration of renewable energy in the cloud-to-fog
architecture has gained recent attention. In \cite{Cui2019}, the authors
propose a solution, based on the Lyapunov Optimization
Technique, that adopts the concept of a request dispatching controller in which all requests are dispatched to. The controller then dispatches requests to nodes that meet the
minimum service time and are powered by sufficient green-
energy, powered by solar cells. In \cite{Lawey2015}, the authors develop an
MILP model, where the effects of wind farms on the locations
of clouds and content replication of the clouds are studied. The
model identifies the number and locations of clouds given the
current number of requests, the transport network power
consumption incurred by having user requests processed at the
new cloud location, and transmission loss of delivering
renewable energy from wind farms. In \cite{Halim2019}, the authors aim to
deliver Video-on-Demand (VoD) traffic by minimizing the
energy reliance on non-renewable energy sources and
maximizing the usage of solar power available to nodes in the access network. Due to the fluctuating nature of solar power
energy, energy storage devices (ESD) were utilized to store
surplus energy. For the core network, this paper considers IP over WDM bypass and Mixed Line Rate.

\section{The Proposed System}

This paper presents an extension to \cite{Alharbi2019}, where the authors develop an MILP
model to place virtualized machine (VM) services in different
nodes in the cloud-fog architecture for optimum energy
efficiency given several factors. These include: the minimum
VM minimum workload requirement to operate, the VM
popularity, and data rate incurred from traffic. The authors
conducted the study on an AT\&T cloud fog architecture with
fixed traffic rates. This paper demonstrates
the impact of placing renewable energy sources, such as solar
power, in the access network on the placement of VMs on
cloud to fog nodes. We consider the dynamic VM
placement given the daily traffic quantities of the BT 21CN
network topology shown in Figure 4. We consider the bypass
approach of IP over WDM whereas \cite{Shen2009} consider the non-
bypass approach. Related research such that of \cite{Halim2019}, acknowledges
that solar power energy is heavily reliant on the current
forecast and the time of the day, therefore, surplus energy
could be stored in either energy storage devices or on the
power grid. This work will not consider ESDs due to limited
battery shelf life and the process of making ESDs has
environmental implications. Although lithium oxides can be
recycled, resource depletion and ecotoxicity arise from
materials such as copper, cobalt, and silver \cite{Kim2018}.

\subsection{Cloud-to-fog Network Design}

The Cloud-to-fog network is composed of multiple layers, as
indicated in Figure 2, the top is the core network,
traversing down to the metro network, and finally to the
access network. The core network is composed of the IP
nodes over the WDM network, which facilitates the communication
throughout the topology. It has two layers, the IP layer,
containing core IP routers that provide maximum bandwidth
to optimize routing to destination and group data packets
based on common traits obtained from low-end edge routers,
and an optical layer. Each core IP router is connected to the 
optical layer that consists of optical switches, which connects to an optical cross
connect, which divides shares the bandwidth into multiple
wavelengths of different frequencies, used in Frequency
division multiplexing, via an associated transponder, that
sends fiber data using a single fiber optic. As the network
topology spans over a long distance, Erbium-Doped Fiber Amplifiers (EDFAs) are used as
amplifiers, this is illustrated by Figure 3. \cite{Osman2016} A
transponder’s main job in IP over WDM is to convert the
the signal received from an IP router into the desired WDM wavelength, and vice-versa at the demultiplexer end. A core
network can be composed of several points of presence.

We design the nodes to contain a limited
amount of IP routers at a given node and have a maximum
amount of wavelengths to a single optical fiber cord. Power
consumption that is utilized from the IP routers, optical
switches, optical cross-connects, transponders, and EDFAs are
summed as part of the objective function, which aims to minimize the
total energy consumption stemming from the components. The
model ensures that the flow conservation law is satisfied such
that the total outgoing flow of data packets is equal to the total
incoming flow if the node, or the point of presence, is an
intermediate node. \cite{Bressan2015} If it is a source node, then the total
output flow subtracted by the total input flow must be equal to
the traffic demand between the node pair. \cite{Bressan2015} If it is a
destination node then the total incoming flow minus the output
flow is equal to the traffic demand between the node pair. This
law ensures that data packets can be diverged throughout
multiple nodes, in both the virtual and physical topologies.
The virtual topology consists of a set of light paths that sits on top
of the physical layer, this ensures that if a physical node is
malfunctioning, then packets are still able to be routed
appropriately. The model ensures that sufficient wavelength
capacity is available in both the physical and virtual
topologies.

\begin{figure}
\label{fig:enter-label}
\centering
    \includegraphics[width=0.85\linewidth]{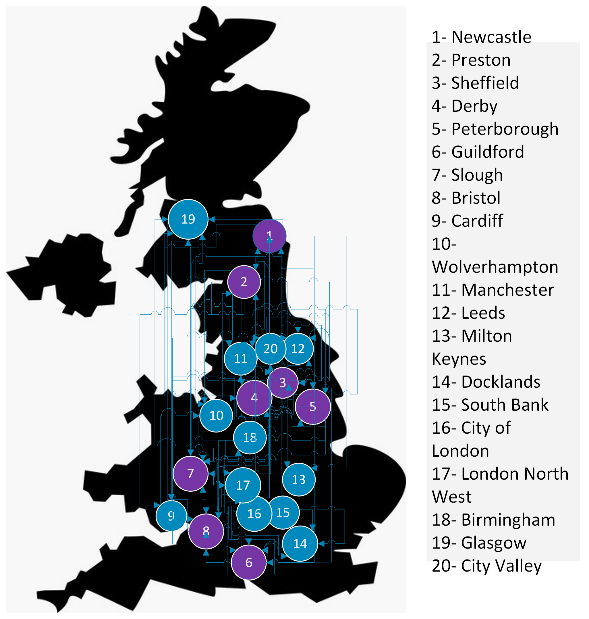}
    \caption{BT 21 CN Network Topology}
\cite{Alharbi20181}
    \cite{Kitz21CN}
\end{figure}

A metro network serves as an aggregation layer that
connects access network node traffic to the core network. It
is consistent of ethernet switches and edge routers. Metro
switches are connected to multiple routers to increase
redundancy.

The access network connects traffic to backbone network,
it is considered the first point of access in which users access
via an ONU, where data is forwarded and received by VMs
directly. In this work, the GPON, which allows for VoIP, IP,
etc. services is the chosen passive-optical network (PON)
standard where the optical line terminal (OLT), connected
by a single optical fiber, acts as the ISP networking endpoint.
This work considers only upstream flow of data, as a downlink
traffic is negligent. It is worth noting that other access
network technologies such as Asymmetric digital
subscriber line (ASDL), which has limited frequency bands,
and Fiber to Node (FTTN) \cite{Cobo2012} \cite{Baliga2009} which incorporates
copper and fiber optic, do exist, however will not be
considered as GPON is adopted for the BT 21CN.

It is worth noting that due to the nature of many modern
applications, users are not constantly imposing workloads to
the network at the access rate stated by the ISP, which means,
user time segmented upload workload is received by the
network by Time Division Multiplexer. We
consider traffic to be imposed only at access rates advertised,
and not at variable access rates throughout the day.

\subsection{Renewable Energy in the Fog Nodes}

This work considers solar power as renewable energy as their
cell sizes can be configured to be appropriately placed in
central offices, located in the PON network (access), near
traffic-producing sources. The available solar power is in units
of W/m2 and the values presented in Table 2 have been
obtained from the SOLPOS calculator based on the BT network
node locations.

\subsection{Datacenter design}

Datacenter servers tend to consume up to 70 percent when in an idle state
\cite{Barroso2007}. In order to combat this, VM consolidation is a technique
that involves clustering VMs to servers with computing
capacity left, however this decreases the performance of the
VM. In this work, VM consolidation is not considered, despite
its overall power efficiency improvement \cite{Kumar2021}. In order to
maintain QoS, a VM service maintains a base workload load,
this occurs from the CPU utilization used to host the OS, in
order to provide the IaaS, SaaS, and PaaS services to users that
request it. In addition, This workload consists of maintaining the memory footprint and listening to incoming requests from
users. In the industry, VM replicas are used to increase service
reliability \cite{Machida2010} to serve users in nodes, the creation of multiple
replicas is due to the fact that the number of users exceeds the
maximum capacity a single VM unit can serve. The MILP
model would determine the number of replicas needed for a
single data center by dividing the total traffic imposed on the
VM in the data center and divide it by the maximum number
of users and their download rate a single VM can serve. As the
number of users increases, the underlying CPU utilization
increases in a linear fashion. This is accounted in the MILP
model.

\begin{table*}
    \caption*{\textbf{Variables}}
    \begin{minipage}{.5\textwidth}
        \centering
        $$
        \begin{tabularx}{\textwidth}{p{1.5cm} X}
        \(W_{jit}^P\) & Number of used wavelengths on physical link between nodes \((j, i)\) at time \(t\) \\
        \(F_{mnt}^P\) & Number of fibers on physical link between nodes \((m,n)\) at time \(t\) \\
        \(W_{jit}^V\) & Wavelengths on the virtual link from nodes \(j,i\) at time \(t\) \\
        \(PT_{ijt}^{sd}\) & Number of wavelength channels between \((s, d)\) through optical layer physical link \(i,j\) at time \(t\) \\
        \(W_{ijt}^V\) & Number of wavelengths in the virtual link between the node pair \(i\) and \(j\) at time \(t\) \\
        \(VT_{ijt}^{sd}\) & Packet traffic flow between node pairs \(s, d\) through virtual link \(i,j\) at time \(t\) \\
        \(F_{ijt}^P\) & Number of fibers on virtual link between nodes \((i,j)\) at time \(t\) \\
        \(W_{ijt}^P\) & Number Wavelengths carried by fiber, between nodes \(i,j\) at time \(t\) \\
        \(C_z^{RA}\) & Number of router ports in node \(z\) that aggregate traffic \\
        \(C_z^{AS}\) & Number of servers at node \(z\) \\
        \(C_z^{SA}\) & Number of switches in cloud node \(z\) \\
        \(M_z^{RA}\) & Number of router ports in fog data center connected to core node \(z\) \\
        \(M_z^{AS}\) & Number of metro servers at node \(z\) in the processing server \\
        \(M_z^{SA}\) & Number of switches in fog data center connected to core node \(z\) \\
        \(MN_z^{AR}\) & Number of metro node routers connected to core node \(z\) \\
        \(MN_z^{ASW}\) & Number of metro node switches connected to core node \(z\) \\
        \(A_z^{RA}\) & Number of access router ports in fog data center connected to core node \(z\) \\
        \(A_z^{AS}\) & Number of access processing servers at node \(z\) in data center \\
        \(A_z^{ASW}\) & Number of access node switches at node \(z\) in data center \\
        \(OLT_{iz}^A\) & Number of OLTs present in PON network \(i\) \\
        \(ONU_{iz}^A\) & Number of ONUs connected to an OLT that is present in PON network \(i\) connected to node \(z\) \\
        \end{tabularx}
        $$
    \end{minipage}%
    \begin{minipage}{.5\textwidth}
        \centering
        $$
        \begin{tabularx}{\textwidth}{p{1.5cm} X}
        \(RVM_{xist}^L\) & Response traffic from the replicated VM \(x\) unit hosted in the data center in PON \(i\) (if access node), connected to core node \(s\), at time \(t\) \\
        \(L_{ixst}^{VM}\) & Binary variable: 1 if a VM replica \(x\) is hosted in access data center, in PON network \(i\), connected to core node \(s\) at time \(t\), else 0 \\
        \(L_{s,p}^{DC}\) & Binary variable in which \(n_{s,p}^{DC} = 1\), if a data center is hosted in network \(L\), connected to core node \(s\), at PON \(i\) (if \(L = \)Access Node) \\
        \(L_{ixst}^{VM}\) & Binary variable in which \(A_{ixst}^{VM} = 1\) if a VM replica \(x\) is hosted in access data center, in PON network \(i\), connected to core node \(s\) at time \(t\), else \(A_{ixst}^{VM} = 0\) \\
        \(VM_{xist}^{TWL}\) & Total workload by summing all VM replica workloads in node \(s\), at PON \(i\) (if \(L = A\)), at time \(t\) \\
        \(L_{is}^{AS}\) & Number of servers in data center connected to PON \(i\) if(\(L = \text{Access Node}\)), connected to core node \(s\) \\
        \(RVM_{xist}^L\) & Traffic flow from VM replica \(x\), hosted on node \(s\) to users at time \(t\) \\
        \(L_s^{AR}\) & Number of routers to be installed in data center in layer \(L\), connected to node \(s\) \\
        \(L_s^{ASW}\) & Number of switches to be installed in data center in layer \(L\), connected to node \(s\) \\
        \(A_{z}^{RBA}\) & Number of access router ports in fog data center connected to core node \(z\) powered by brown sources \\
        \(A_{z}^{ABS}\) & Number of access processing servers at node \(z\) in data center powered by brown power sources \\
        \(A_{is}^{AS}\) & Total number of servers, derived from the number of both brown and green powered \\
        \(A_{is}^{ASW}\) & Total number of switches, derived from the number of both brown and green powered \\
        \(A_{is}^{AR}\) & Total number of routers, derived from the number of both brown and green powered \\
        \(A_{is}^{AGS}\) & Number of green powered servers at node \(s\), at PON \(i\) \\
        \(A_{is}^{AGSW}\) & Number of green switches servers at node \(s\), at PON \(i\) \\
        \(A_{is}^{AGR}\) & Number of green-powered routers at node \(s\), at PON \(i\) \\
        \(RVM_{xist}^A\) & Total response traffic from all VMs at PON \(i\) connected at node \(s\) at time \(t\) \\
        \end{tabularx}
        $$
    \end{minipage}
\end{table*}

\begin{table*}
    \centering
    \caption*{\textbf{Parameters}}
    \begin{minipage}{.5\textwidth}
        \centering
        $$
        \begin{tabularx}{\textwidth}{p{1.5cm} X}
        \(C^{PUE}\) & Power Usage Effectiveness of Cloud \\
        \(S\) & Distance between EDFAs (KM) \\
        \(D_{ij}\) & Physical link distance, between the nodes in N (KM) \\
        \(T\) & Time of day (HR) \\
        \(N\) & Set of IP over WDM nodes \\
        \(A_{mn}\) & \(D_{mn}/ (S /- 1) + 2\). Number of EDFAs on the physical link between \( (m, n) \) from \(N\) \\
        \(EDFA^P\) & EDFA Power Consumption (W) \\
        \(T^P\) & Power Consumption of a Transponder (W) \\
        \(RP^{P}\) & Router Port Power Consumption (W) \\
        \(RPs\) & Number of router ports that aggregate traffic from metro routers at node \(s\) \\
        \(T_t^{sd}\) & Traffic between node pair \(s, d\) at time \(t\) \\
        \(WN\) & Wavelength data rate or capacity (Gb/s) \\
        \(C^{RP}\) & Cloud Fog Router Power Consumption (W) \\
        \(C^{SWP}\) & Cloud Switch Power Consumption (W) \\
        \(CF^R\) & Switch Redundancy cloud and fog \\
        \(C^{PUE}\) & Cloud Data Center Power Usage Effectiveness \\
        \(SP\) & Server Power Consumption (W) \\
        \(P_{CN}\) & Power consumption of core network (W) \\
        \(P_{CS}\) & Power consumption of cloud servers (W) \\
        \(P_{CRS}\) & Power consumption of cloud routers and switches (W) \\        
        \end{tabularx}
        $$
    \end{minipage}%
    \begin{minipage}{.5\textwidth}
        \centering
        $$
        \begin{tabularx}{\textwidth}{p{1.5cm} X}
        \(M^{RP}\) & Metro Router Power Consumption (W) \\
        \(M^{SWP}\) & Metro Switch Power Consumption \\
        \(MF^R\) & Metro Fog Redundancy \\
        \(MN^{RP}\) & Metro Node Router Power Consumption \\
        \(MN^{SWP}\) & Metro Node Switch Power Consumption \\
        \(MN^R\) & Metro Node Redundancy \\
        \(M^{PUE}\) & Metro Data Center Power Usage Effectiveness \\
        \(A^{RP}\) & Access fog router power consumption \\
        \(A^{SWP}\) & Access fog switch power consumption \\
        \(OLT^P\) & OLT power Consumption \\
        \(ONU^P\) & ONU power consumption \\
        \(A^{PUE}\) & Access Data Center Power Usage Effectiveness \\
        \(AF^R\) & Access Data Center Switch \\
        \(RT_{xidt}\) & Traffic flow from VM \(x\) to users in PON \(i\) connected to core node \(d\), at time \(t\) \\
        \(OLT_{iz}\) & OLT Capacity at PON \(i\), connected to node \(z\) \\
        \(A_i\) & Broadband rate \\
        \(USER_{xidt}\) & Number of users at PON \(i\), connected to node \(d\) accessing VM replica \(x\), at time \(t\) \\
        \(USER_x^D\) & User download rate \\
        \end{tabularx}
        $$
    \end{minipage}
    \begin{minipage}{.5\textwidth}
        \centering
        $$
        \begin{tabularx}{\textwidth}{p{1.5cm} X}
        \(Z\) & Very Large Number \\
        \(RT_{xidt}\) & \(\frac{\text{Number of users at PON } i \times \text{User Data Rate}}{VM_x^{maxW} - VM_x^{minW}} \times USER_x^{MD}\) \\
        \(VM_x^{maxW}\) & Maximum workload of VM \(x\) \\
        \(VM_x^{minW}\) & Minimum workload requirement for VM \(x\). This equation represents workload per traffic unit \\
        \end{tabularx}
        $$
    \end{minipage}%
    \begin{minipage}{.5\textwidth}
        \centering
        $$
        \begin{tabularx}{\textwidth}{p{1.5cm} X}
        \(USER_x^D\) & User data rate accessing VM \(x\) \\
        \(USER_x^m\) & Maximum number of users VM \(x\) can support \\
        \(M\) & Minimum workload requirement \\
        \end{tabularx}
        $$
    \end{minipage}
\end{table*}

\vspace*{50pt}

\section{MILP Model}

\begin{center}
\textbf{Objective Function: Minimize}
\end{center}

The objective of the MILP model is to minimize the total energy consumption, derived from non-renewable sources (grid), therefore the model will reduce the amount of servers, routers, switches, ONUs, and OLTs used in the network

\begin{equation}
P_{CN} = P_{\text{transponder}} + P_{\text{EDFA}} + P_{\text{routerPort}}
\end{equation}

\begin{equation}
P_{transponder} = \Sigma_{j\epsilon N} \Sigma_{i \epsilon N_m} W_{jit} T^{p}(t)
\end{equation}

\begin{equation}
P_{\text{EDFA}} = \sum_{t \in T}\sum_{j \in N} \sum_{i \in N_m} EDFA^P \left( {\frac{D_{ij}}{S-1} + {2}}F^P_{mnt} \right)
\end{equation}

\begin{equation}
P_{\text{routerPort}} = \sum_{j \in N} RP^{P} \left( RPs + \sum_{t \in T} \sum_{\substack{i \in N \\ j \neq i}} W^V_{jit} \right)
\end{equation}

\begin{equation}
PWR = P_{Metro} + P_{Core} + P_{Access}
\end{equation}
\begin{equation}
\begin{split}
P_{Core} = & \ P_{CS} \\
        & + P_{CSR} \\
        & + P_{CN}
\end{split}
\end{equation}

\begin{equation}
P_{CSR} = \sum_{t \in T} \sum_{z \in N} C^{\text{RP}}C_z^{\text{RA}} + (C^{\text{SWP}}C_z^{\text{SA}})
\end{equation}

\begin{equation}
P_{CS} = C^{\text{PUE}} \sum_{t \in T} \sum_{z \in N} C_z^{\text{AS}}S^P
\end{equation}

\begin{equation}
\text{CoreNetwork} = P_{\text{transponder}} + P_{\text{EDFA}} + P_{\text{routerPort}}
\end{equation}

\begin{equation}
    \text{MetroP}_{\text{datacenter}} = \text{MetroP}_{\text{server}} + \text{MetroP}_{\text{datacenterRouterSwitches}}
\end{equation}

\begin{equation}
    \text{MetroP}_{\text{server}} = M^{\text{PUE}} \sum_{z \in N} M_z^{\text{AS}} S^P
\end{equation}

\begin{IEEEeqnarray}{rCl}
    \text{MetroP}_{\text{datacenterRouterSwitches}} &=& \sum_{t \in T} \sum_{z \in N} \left( M^{\text{RP}}M_z^{\text{RA}} \right) \nonumber\\
    && \quad  + M^{\text{SWP}}MF^{\text{R}}M_z^{\text{SA}}
\end{IEEEeqnarray}

\begin{equation}
    \text{MetroP}_{\text{router}} = \sum_{z \in N} \text{MN}^{\text{RP}}\text{MN}^R{\text{MN}}_z^{\text{AR}}
\end{equation}

\begin{equation}
    \text{MetroP}_{\text{switch}} = \sum_{z \in N} \text{MN}^{\text{SWP}}\text{MN}_z^{\text{ASW}}
\end{equation}

\begin{equation}
    \text{AccessP} = \text{AccessP}_{\text{datacenter}} + P^{\text{ONU}} + P^{\text{OLT}}
\end{equation}

\begin{equation}
\text{AccessP}_{\text{datacenter}} = \text{AccessP}_{\text{server}} +\text{AccessP}_{\text{datacenterRouterSwitches}}
\end{equation}

\begin{equation}
    \text{AccessP}_{\text{server}} = A_{\text{PUE}} \sum_{z \in N} A_z^{\text{BASS}} P
\end{equation}

\begin{IEEEeqnarray}{rCl}
    \text{AccessP}_{\text{datacenterRouterSwitches}} &=& \sum_{t \in T} \sum_{z \in N} \Big( A^{\text{RP}}A_z^{\text{BRA}} \nonumber\\
    && +\> A^{\text{SWP}}AF^{\text{R}}A_z^{\text{BSA}} \Big)
\end{IEEEeqnarray}

\begin{equation}
    P_{OLT} = \sum_{i \in P} \sum_{z \in N} OLT^P (OLT_{iz}^A)
\end{equation}

\begin{equation}
    P_{ONU} = \sum_{i \in P} \sum_{z \in N} ONU^P (ONU_{iz}^A)
\end{equation}

\begin{center}
\textbf{Constraints}
\end{center}

\begin{equation}
    \sum_{s \in N} \sum_{d \in N} P_{ijt}^{sd} \leq W_{ijt}^P F_{ijt}^P, \quad i \in N, j \in N_m, t \in T: i \neq j
\end{equation}
Constraint (21) Maintains traffic quantity under physical link capacity

\begin{equation}
    \sum_{s \in N} \sum_{d \in N} V_{ijt}^{sd} \leq W_{ijt}^V WN, \quad i,j \in N, t \in T: s \neq d
\end{equation}

Constraint (22) Maintains traffic quantity under virtual link capacity

\begin{equation}
\begin{split}
    \sum_{\substack{i \in N \\ i \neq j}} V_{jit}^{sd} - \sum_{\substack{i \in N \\ i \neq j}} V_{ijt}^{sd} &= 
    \left\{
    \begin{array}{ll}
        T_{t}^{sd} & \text{if } m = s \\
        -T_{t}^{sd} & \text{if } m = d \\
        0 & \text{otherwise}
    \end{array}   
    \right. \\
    & \quad \forall s, d, j \in N, \> \forall t \in T, \> s \neq d
\end{split}
\end{equation}

Constraint (23) maintains the flow conservation law in the core
network virtual link. It ensures that packets are directed from the source to the destination. If the router is an intermediate router, then the incoming traffic is equal to the outgoing traffic

\begin{equation}
\begin{split}
    \sum_{j \in N_m} PT_{ijt}^{sd} - \sum_{j \in N_m} PT_{jit}^{sd} &= 
    \left\{
    \begin{array}{ll}
        W_{ijt}^V & \text{if } i = s \\
        -W_{ijt}^V & \text{if } i = d \\
        0 & \text{else}
    \end{array} 
    \right.\\
    & \quad \forall t \in T, \; \forall s, d, i \in N, \; i \neq d
\end{split}
\end{equation}

Constraint (24) Maintains the physical flow conservation law in which incoming multiplexed wavelengths are either absorbed or sent.

\begin{IEEEeqnarray}{rCl}
    \sum_{i \in P} \sum_{d \in N} RT_{xidt} &=& \sum_{s \in N} \sum_{i \in P} RVM_{xist}^A + \sum_{s \in N} RVM_{xst}^M \nonumber\\
    && +\> \sum_{s \in N} \sum_{d \in N} RVM_{xsdt}^C, \nonumber\\
    && \quad \forall x \in VM, \> \forall t \in T
\end{IEEEeqnarray}

Constraint (25) ensures that the demand from traffic-producing sources is satisfied with a VM in a data center in a core network and/or a replica VM in a data center in a metro and PON networks

\begin{equation}
\begin{split}
    RT_{xidt} &= USER_{xidt} \cdot USER_{x}^D, \\
    & \quad \forall x \in VM, \> \forall t \in T, \> \forall d \in N, \> \forall i \in P
\end{split}
\end{equation}

Constraint (26) calculates the maximum bit rate imposed by the maximum amount of users a single VM replica can handle

\begin{equation}
\begin{split}
    RVM_{xist}^A &\geq A_{ixst}^{VM}, \\
    & \forall i \in P, \> \forall t \in T, \> \forall s \in N, \> \forall x \in VM
\end{split}
\end{equation}

\begin{equation}
\begin{split}
    RVM_{xist}^A &\leq A_{ixst}^{VM} Z, \\
    & \forall i \in P, \> \forall t \in T, \> \forall s \in N, \> \forall x \in VM
\end{split}
\end{equation}

\begin{equation}
\begin{split}
    RVM_{xst}^M &\geq M_{xst}^{VM}, \\
    & \forall t \in T, \> \forall s \in N, \> \forall x \in VM
\end{split}
\end{equation}

\begin{equation}
\begin{split}
    RVM_{xst}^M &\leq M_{xst}^{VM} Z, \\
    & \forall t \in T, \> \forall s \in N, \> \forall x \in VM
\end{split}
\end{equation}

\begin{equation}
\begin{split}
    Z \sum_{d \in N} RVM_{xsdt}^C &\geq C_{xst}^{VM}, \\
    & \text{for } \forall x \in VM, \> \forall t \in T, \> \forall s \in N
\end{split}
\end{equation}

\begin{equation}
\begin{split}
    \sum_{d \in N} RVM_{xsdt}^C &\leq C_{xst}^{VM} Z, \\
    & \forall t \in T, \> \forall s \in N, \> \forall x \in VM
\end{split}
\end{equation}

Constraints (27-32) simulate the placement of a VM in a metro, PON and core networks

\begin{equation}
    \sum_{x \in VM} A_{ixs}^{VM} \geq A_{si}^{DC}, \quad s \in N, \quad i \in P
\end{equation}

\begin{equation}
    \sum_{x \in VM} A_{ixs}^{VM} \leq A_{si}^{DC} Z, \quad s \in N, \quad i \in P
\end{equation}

  \begin{equation}
    \sum_{x \in VM} M_{xs}^{VM} \geq M_{s}^{DC}, \quad s \in N, \quad i \in P
\end{equation}

\begin{equation}
    \sum_{x \in VM} M_{xs}^{VM} \leq M_{s}^{DC} Z, \quad s \in N, \quad i \in P
\end{equation}

\begin{equation}
    \sum_{x \in VM} C_{xs}^{VM} \geq C_{s}^{DC}, \quad s \in N, \quad i \in P
\end{equation}

\begin{equation}
    \sum_{x \in VM} C_{xs}^{VM} \leq C_{s}^{DC} Z, \quad s \in N, \quad i \in P
\end{equation}

Constraints (33-38) simulate the placement of a datacenters, that are able to host VM services, in the core, metro, and PON nodes.

\begin{equation}
    \text{USER}_{x}^{MD} = \text{USER}_{x}^{D} \text{USER}_{x}^{m}, \quad x \in VM
\end{equation}

Constraint (39) calculates the maximum bit rate imposed by the maximum amount of users a single VM replica can handle

\begin{IEEEeqnarray}{rCl}
    A_{xst}^{TVM} &=& \frac{VM_{x}^{maxW} - VM_{x}^{minW}}{USER_{x}^{MD} \cdot RVM_{xst}^{A}} + A_{xst}^{VM} \nonumber\\
    && \times \left( \frac{RVM_{xsdt}^{A}}{USER_{x}^{MD}} \right), \nonumber\\
    && \quad \forall t \in T, \> \forall x \in VM, \> \forall s \in N, \> \forall i \in P
\end{IEEEeqnarray}

\begin{IEEEeqnarray}{rCl}
    M_{xst}^{TVM} &=& \frac{(VM_{x}^{maxW} - VM_{x}^{minW})RVM_{xst}^{M}}{USER_{x}^{MD}} \nonumber\\
    && +\> M_{xst}^{VM} M  \frac{RVM_{xsdt}^{M}}{USER_{x}^{MD}}, \nonumber\\
    && \quad \forall t \in T, \> \forall x \in VM, \> \forall s \in N
\end{IEEEeqnarray}

\begin{IEEEeqnarray}{rCl}
    C_{xst}^{TVM} &=& \frac{(VM_{x}^{maxW} - VM_{x}^{minW})}{USER_{x}^{MD}} \sum_{d \in N} {RVM_{xsdt}^{C}} \nonumber\\
    && + C_{xst}^{VM}M \sum_{d \in N} \frac{RVM_{xsdt}^{C}}{USER_{x}^{MD}}, \nonumber\\
    && \quad \forall t \in T, \> \forall x \in VM, \> \forall s \in N, \> \forall i \in P
\end{IEEEeqnarray}

\begin{IEEEeqnarray}{rCl}
    A_{xst}^{TVM} &=& VM_{x}^{maxW}A_{xst}^{VM}, \nonumber\\
    && \quad \forall t \in T, \> \forall x \in VM, \> \forall s \in N
\end{IEEEeqnarray}

\begin{IEEEeqnarray}{rCl}
    M_{xst}^{TVM} &=& VM_{x}^{maxW}M_{xst}^{VM}, \nonumber\\
    && \quad \forall t \in T, \> \forall x \in VM, \> \forall s \in N
\end{IEEEeqnarray}

\begin{IEEEeqnarray}{rCl}
    C_{xst}^{TVM} &=& VM_{x}^{maxW}C_{xst}^{VM}, \nonumber\\
    && \quad \forall t \in T, \> \forall x \in VM, \> \forall s \in N
\end{IEEEeqnarray}

Constraints (40-42) calculate the total workload of a single VM placed in a data center that is hosted in a metro network node, a PON network node, or a core node. These constraints consist of two parts. The first part calculates the linear correlation of CPU utilization based on the number of users based on workload per traffic unit. The second part calculates the number of replicas needed in the data center based on the total workflow imposed on the data center and constraint (40). Constraints (43-45), calculate the workload of a VM under a constant CPU profile

\begin{IEEEeqnarray}{rCl}
    VM_{ist}^{TWA} &=& \sum_{x \in VM} A_{xst}^{TVM}, \nonumber\\
    && \quad \forall t \in T, \> \forall x \in VM, \> \forall s \in N, \> \forall i \in P
\end{IEEEeqnarray}

\begin{equation}
    VM_{st}^{TWM} = \sum_{x \in VM} M_{xst}^{TVM}, \quad t \in T, \quad x \in VM, \quad s \in N
\end{equation}

\begin{IEEEeqnarray}{rCl}
    VM_{st}^{TWC} &=& \sum_{x \in VM} C_{xst}^{TVM}, \nonumber\\
    && \quad \forall t \in T, \> \forall x \in VM, \> \forall s \in N, \> \forall i \in P
\end{IEEEeqnarray}

Constraints (46-48), calculates the total workloads from the traffic by summing all the workloads of the VM replicas hosted in the datacenters located in the access, metro and core networks

\begin{equation}
    \frac{VM_{xizt}^{TWA}}{S^{\text{max}}} \leq A_{is}^{AS}, \quad i \in P, \quad s \in N
\end{equation}

\begin{equation}
    \frac{VM_{xst}^{TWM}}{S^{max}} \leq M_{s}^{AS}, \quad s \in N 
\end{equation}

\begin{equation}
    \frac{VM_{xst}^{TWC}}{S^{max}} \leq C_{s}^{AS}, \quad s \in N
\end{equation}

Constraints (49-51) calculate the number of servers needed each data center of the cloud fog network. The calculations are derived from the total workload imposed on the nodes divided by the maximum workload capacity a server can handle. A VM service utilizes the underlying server’s CPU

\begin{equation}
\begin{split}
    \frac{\sum_{x \in VM} RVM_{xst}^{A}}{A^{RBR}} &\leq A_{is}^{AR}, \\
    & \forall s \in N, \> \forall i \in P, \> \forall t \in T
\end{split}
\end{equation}

\begin{equation}
    \frac{\sum_{s \in N} \sum_{v \in VM} RVM_{xst}^{C}}{C^{RBR}} \leq C_{s}^{AR}, \quad s \in N
\end{equation}

\begin{equation}
    \frac{\sum_{x \in VM} RVM_{xst}^{M}}{M^{RBR}} \leq M_{s}^{AR}, \quad s \in N
\end{equation}

Constraints (52-54) calculate the number of routers needed each data center of the cloud fog network. The calculations are derived from the total traffic imposed on the nodes divided by the bit rate of the respective routers

\begin{equation}
    \frac{\sum_{v \in VM} RVM_{xst}^{A}}{ASWB} \leq A_{is}^{ASW}, \quad s \in N, \quad i \in P, \quad t \in T
\end{equation}

\begin{equation}
    \frac{\sum_{d \in N} \sum_{v \in VM} RVM_{xst}^{C}}{CSWB} \leq C_{s}^{ASW}, \quad s \in N
\end{equation}

\begin{equation}
    \sum_{v \in VM} \frac{RVM_{xst}^{M}}{MSWB} \leq M_{s}^{ASW}, \quad s \in N
\end{equation}

Constraints (55-57) calculate the number of switches needed each data center of the cloud fog network. The calculations are derived from the total traffic imposed on the nodes divided by the bit rate of the respective routers

We now introduce novel constraints to the model to ensure green power is dedicated to routers, switches, and serves in the access fog data center:

\begin{equation}
    A_{is}^{AS} = A_{is}^{AGS} + A_{is}^{BAS}, \quad i \in P, s \in N
\end{equation}

Constraint (58) Indicates the total number of servers in an access fog, in PON network i, connected to node z, is consistent with servers powered by both green and non-renewable energy

\begin{equation}
    A_{\text{is}}^{\text{ASW}} = A_{\text{is}}^{\text{AGSW}} + A_{\text{is}}^{\text{BASW}}, \quad i \in P, s \in N
\end{equation}

Constraint (60) Indicates the total number of routers in an access fog, in the PON network i, connected to node z is consistent of servers powered by both green and non-renewable energy

\begin{equation}
    A_{\text{is}}^{\text{AR}} = A_{\text{is}}^{\text{AGR}} + A_{\text{is}}^{\text{BAR}}, \quad i \in P, s \in N
\end{equation}

Constraint (59) Indicates the total number of switches in an access fog, in PON network i, connected to node n is consistent of servers powered by both green and non-renewable energy

\begin{IEEEeqnarray}{rCl}
    A_{is}^{AGS} &=& \frac{{SP_{ist}^{m} SP - \left( A_{is}^{AGSW}A^{SWP} + A_{is}^{AGR}A^{RP} \right)}}{S^{P}}, \nonumber\\
    && \quad \forall t \in T, i \in P, s \in N
\end{IEEEeqnarray}

Constraint (61) Indicates the number of servers at PON p, connected to node z, powered by the available green energy at time t

\begin{equation}
    A_{is}^{AGS} \leq \frac{VM_{ist}^{TWA}}{S^{\text{max}}} \quad \forall t \in T, i \in P, s \in N
\end{equation}

Constraint (62) Indicates the number of routers at PON p, connected to node z, powered by the available green energy at time t

\begin{IEEEeqnarray}{rCl}
    A^{GR} &=& \frac{{S^{P}_{ist} m^{SP} - \left( A_{is}^{AGS}S^{P} + A^{SWP}A_{is}^{AGSW} \right)}}{{A^{RP}}} \nonumber\\
    && ,\> \forall t \in T, i \in P, s \in N
\end{IEEEeqnarray}

Constraint (63) Indicates the number of switches at PON p, connected to node z, powered by the available green energy at time t

\begin{equation}
    A_{is}^{GR} \leq \frac{{\sum_{x \in VM} RVM_{xist}^A}}{{A^{RBR}}} \, \\, \forall t \in T, i \in P, s \in N
\end{equation}

Constraint (64) Indicates the number of servers at PON p, connected to node z, powered by the available green energy at
time t, needs to be less than the total number of servers

\begin{IEEEeqnarray}{rCl}
    A_{is}^{AGSW} &=& \frac{{SP_{ist}^{m} SP - \left( A_{is}^{AGS} S^P + A^{RP} A_{is}^{AGR} \right)}}{{A^{SWP}}} \nonumber\\
    && ,\> \forall t \in T, i \in P, s \in N
\end{IEEEeqnarray}

Constraint (66) Indicates the number of routers at PON p, connected to node z, powered by the available green energy at time t, needs to be less than the total number of routers $${A_{iz}}$$

\begin{equation}
    A_{is}^{AGSW} \leq \sum_{x \in VM}  \frac{R^{VMA}_{xist}}{A^{SWB}} \, , \forall t \in T, i \in P, s \in N 
\end{equation}

Constraint (67) Indicates the number of switches at PON p, connected to node z, powered by the available green energy at time t, needs to be less than the total number of switches
In which the number of brown-powered servers, switches, and routers placed in the data centers is minimized by the model.

\begin{figure}
    \centering
    \includegraphics[width=1.0\linewidth]{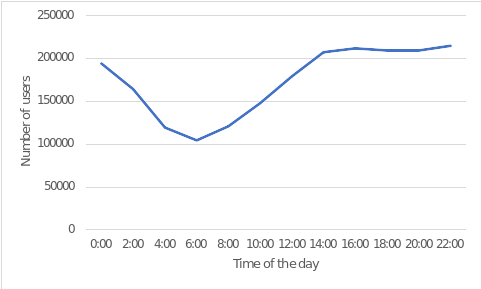}
    \caption{BT 21 CN Average Traffic}
    \label{fig:BT-21-CN}
\end{figure}

\section{Results}

\begin{figure}
    \centering
    \includegraphics[width=1.0\linewidth]{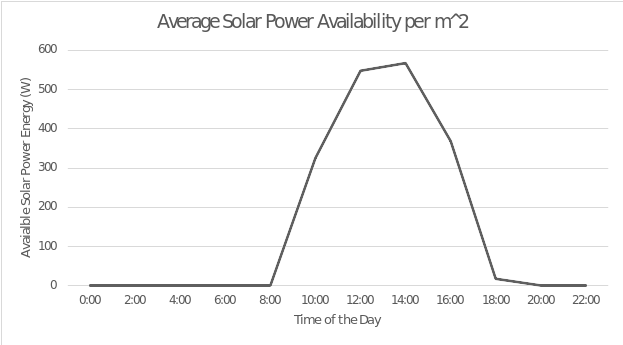}
    \caption{Average Solar Power Availability}
    \label{fig:average-solarl}
\end{figure}
\begin{figure*}[!htbp]
    \centering
    \makebox[\textwidth][c]{%
        \includegraphics[width=1.15\textwidth]{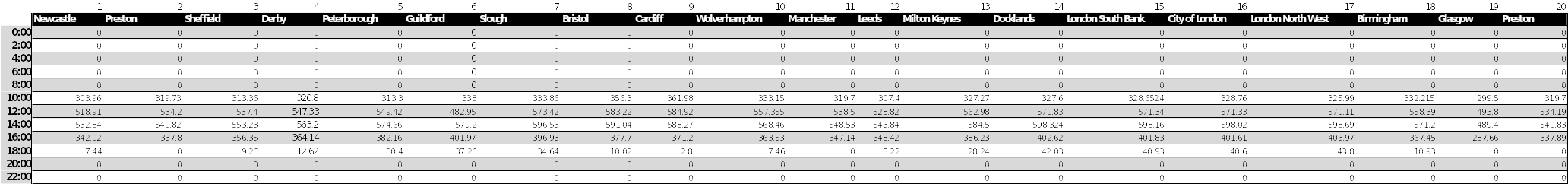}
    }
    \caption{Solar Power Availability at Different Nodes (W)}
    \label{fig:solar-power-availability}
\end{figure*}

\begin{table}[h]
    \caption{Access Fog Nodes and Access Network Parameters}
    \begin{tabular}{p{4cm}p{2cm}}
        \toprule
        \textbf{Parameter} & \textbf{Input Value} \\
        \midrule
        PUE of an Access Node & 1.5 \\
        Access Fog Switch Bit Rate & 160 Gbps \\
        Access Fog Switch Power Consumption & 102W \\
        Access Router Port Bit Rate & 40 Gbps \\
        Access Router Power Consumption & 13 W \\
        \bottomrule
    \end{tabular}
    \label{tab:access_fog_net_params}
\end{table}
\begin{table}[h]
    \caption{Metro Fog Nodes and Metro Network Parameters}
    \begin{tabular}{p{4cm}p{2cm}}
        \toprule
        \textbf{Parameter} & \textbf{Input Value} \\
        \midrule
        PUE of a Metro Node & 1.4 \\
        Metro Fog Switch Bit Rate & 600 Gbps \\
        Metro Fog Switch Power Consumption & 470 W \\
        Metro Fog Router Bit Rate & 600 Gbps \\
        Metro Fog Router Power Consumption & 13W \\
        Metro Network Node Router Bit Rate & 40 GBps \\
        Metro Network Node Power Consumption & 30W \\
        Metro Network Switch Bit Rate & 384 Gbps \\
        Metro Network Switch Power Consumption & 55 W \\
        Metro Fog Router Port Bit rate & 40 Gbps \\
        \bottomrule
    \end{tabular}
    \label{tab:metro_fog_net_params}
\end{table}
\begin{table}[h]
    \caption{Core Network Inputs}
    \begin{tabular}{p{4cm}p{2cm}}
        \toprule
        \textbf{Parameter} & \textbf{Input Value} \\
        \midrule
        Number of Wavelengths per fiber between (m, n) & 32 \\
        Wavelength data rate & 40 GBps \\
        Maximum Distance between two EDFAs in km & 80 km \\
        Power consumption of an EDFA & 55 W \\
        Power consumption of a Transponder & 167 W \\
        Power Consumption of an IP router in the core network & 638 W \\
        PUE of a Cloud Node & 1.3 \\
        Cloud Fog Switch Bit Rate & 600 Gbps \\
        Cloud Fog Switch Power Consumption & 470 W \\
        Cloud Fog Router Bit Rate & 40 GBps \\
        Cloud Fog Router Power Consumption & 30W \\
        Cloud router Port Bit Rate & 40Gbps \\
        \bottomrule
    \end{tabular}
    \label{tab:core_net_params}
\end{table}
\begin{table}[h]
    \caption{General Parameters}
    \begin{tabular}{p{4cm}p{2cm}}
        \toprule
        \textbf{Parameter} & \textbf{Input Value} \\
        \midrule
        Server Power Consumption & 450 W \\
        Maximum Server Workload & 100\% \\
        Metro Network Node Router Redundancy & 2 \\
        \bottomrule
    \end{tabular}
\end{table}
\begin{table}[h]
    \caption{VM Input}
    \begin{tabular}{p{4cm}p{2cm}}
        \toprule
        \textbf{Parameter} & \textbf{Input Value} \\
        \midrule
        Number of Virtual Machines & 1 \\
        User download rate from accessing VM x & \{2 MBps\} \\
        Maximum number of users VM x can serve & 640 \\
        Percentage of underlying CPU that VM x utilizes for its maximum workload capacity & \{2\%, 50\%\} \\
        Minimum workload of VM x when no traffic is imposed & 30\% \\
        \bottomrule
    \end{tabular}
\end{table}
\begin{table}[h]
    \caption{PON Network Input}
    \begin{tabular}{p{4cm}p{2cm}}
        \toprule
        \textbf{Parameter} & \textbf{Input Value} \\
        \midrule
        Power Consumption of an OLT & 1842 W \\
        OLT Capacity located in PON i linked to node z & 1280 Gbps \\
        Number of OLTs present in PON network p & 1 \\
        Power consumption of an ONU & 5 W \\
        \bottomrule
    \end{tabular}
\end{table}

In the cloud-to-fog, the power consumption of networking
components varies, where components such as switches in the
cloud and core networks consume greater power, and switches
in the access networks tend to consume less power. The BT 21 CN network topology consists of 20 nodes and 68 bidirectional links as demonstrated in Figure 3. A node is an IP core router that supports multiple routing protocols and is connected via an optical switch to other nodes in the backbone. A node connects to
the metro network via edge routers connected to the broadband network gateway. The access network is accessed via an
ethernet switch connected to an OLT, which supports
ONUs that allow users to access the network. Figure 4 indicates the varying traffic quantity over a day period obtained in
2016 \cite{Alharbi2018}. According to \cite{NetworkWorldBT}, a list of networking vendors provide networking equipment for the BT
21 CN. This includes Cisco and Lucent for the backbone core
nodes. Alcatel, Siemens, and Alcatel for the metro switches
and edge routers. Ciena for the optical fibers that allow for the
IP over WDM. Fujitsu for access network supplier. As the BT
21 CN topology utilizes the Dense-Wavelength Division
Multiplexing (DWDM) the backbone, the Cisco CRS-1
Carrier Routing System is the adopted core IP router that
provides for 40 Gbps throughput at 638 W of power, that
provides routing of voice, data, and video induced to VMs.The Cisco IOS CR7 \cite{CiscoNCS5500} is the adopted network operating
system, therefore, the edge router located in the metro network
is routed via the Cisco NCS-5502-SE Chassis with 30 W
power consumption per 40 Gbps router port and is connected
to the CRS-1 via a packet over SONET 10 Gbps link. The
Cisco NCS-5502 is the router adopted in the data centers located
in the metro and access networks, where 30 W of power is
consumed per 40 Gbps. The Alcatel-Lucent OmniSwitch 6450
is the adopted switch for the data centers which consume 102
W of power and has a bit rate of 162 Gbps. This work
considers tree-and-branch PON topology to connect users to the network. An optical fiber, provided by Ciena, connects to
a splitter that splits the fiber link to time-shared user ONUs.
The Wave7 ONT-G1000i is the adopted ONU, which
consumes 5W of power, and communicates with the Hitachi
1220 OLT device, which consumes 1842 W of power. The
MILP models in this paper have been solved on an AMD
Ryzen 7 4800H with 8 Core CPUs. 

Figure 8 illustrates the number of wavelengths needed in the physical
and virtual links in order to fulfill traffic. Figure 7
indicates a linear increase in power consumption with
increased traffic demand. The results show that IP routers consume almost 90 percent of the total power consumption, whilst EDFAs, Optical Cross Connects, etc. consume a smaller portion in the core node. This is due to the fact that IP routers are directly
managing user traffic with multiple routing algorithms \cite{VanHeddeghem2012}.

We investigate the optimal VM placement, given different workloads and VM workload profiles throughout the day at time t, with renewable energy placed in the
access data centers.  We do not consider VM replication as the primary reason for increasing
redundancy and acknowledge that replicas throughout the
network may not be consistent. The results indicate that the VM placement varies due to the amount of traffic.

 Three workload profiles: 25 percent, 50 percent,
and 100 percent of the underlying CPU capacity of the host server,
workload profile (constant or linear), and user data rates of 2
Mbps, which for the most part satisfies basic user tasks
such as social media, text messaging, etc. Ideal PUE values of 1.5, 1.4,
and 1.3 for the core, metro, and access networks where utilized
\cite{Zoie2017}.

Figures 12,13 show Under the constant workload, it was indicated that as the
workload of VMs decreases, from decreased traffic demand, more VM replication occurs in the
metro and access fogs. This is due to the fact that computing
resources in the fog layers are limited, which is sufficient for VMs
with small workloads. This is the opposite case for periods of increased traffic where VMs are placed in the clouds, whereas the VMs are allowed more computational resources. 

Under the linear workload, it has also been shown that workload is proportional
to traffic, Figures 9, 10, and 11
demonstrate the VM placement in which the user data is 2
Mbps, and indicate the same behavior, however, lesser
offloading to access and metro layers.
The total power consumption, under a linear workload, with
a minimum of 2 percent CPU utilization profile (to maintain QoS,
and OS, etc.). The figure indicates that user
traffic is proportional to power overall consumption.

It is clear that whenever solar
power was high, between 12:00 – 16:00, VM replicas tended
to be placed on the access nodes. This is due to the reason that
the overall PUE of the access data center improved. In addition,
transport cost is saved with the addition of solar power. Under
constant workload, VMs were placed primarily in clouds
whenever solar power was not available and were placed in
access data centers whenever solar power was available.
VM replicas are placed towards the access
fog data centers whenever traffic and user data rate is
increased. This means that the MILP accounted for the fact
that an increased number of users requesting core data centers
is less efficient.
Under the linear workload, VM replicas tended to be placed
in the access nodes as VM workload decreased. Whenever
traffic was high, VMs were placed in the cloud and metro
nodes.
Overall power consumption was reduced by 16 percent as indicated in Figure 15 compared to Figure 14, where no solar power was noticed, this is noticeable in times 10:00 – 14:00, where solar power was high, and user
the data rate was relatively low.

\begin{figure}
    \centering
    \includegraphics[width=1.0\linewidth]{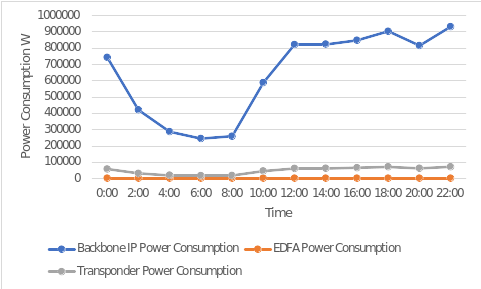}
    \caption{Power consumption of various core node components due to varying daily traffic}
\end{figure}

\begin{figure}
    \centering
    \includegraphics[width=1.0\linewidth]{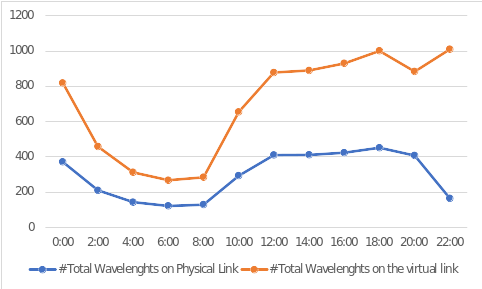}
    \caption{Number of wavelengths a physical and virtual link utilized due to varying daily traffic}
\end{figure}

\begin{figure}
    \centering
    \includegraphics[width=1.0\linewidth]{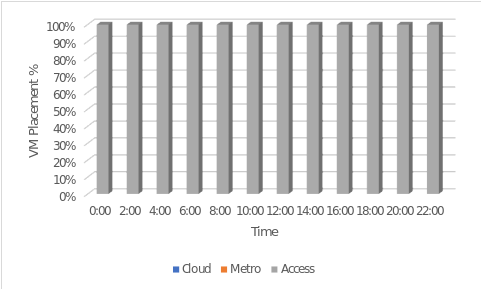}
    \caption{VM Placement in Cloud Fog under linear workload, 25 percent VM workload, and user download rate of 2 Mbps}
\end{figure}

\begin{figure}
    \centering
    \includegraphics[width=1.0\linewidth]{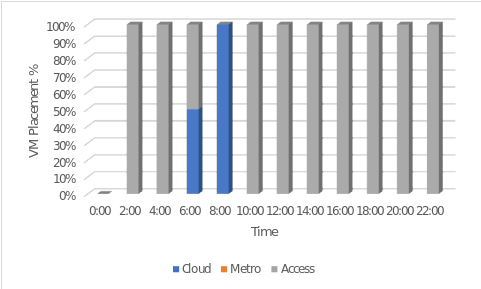}
    \caption{VM Placement in Cloud Fog under linear workload, 50 percent VM workload, and user download rate of 2 Mbps}
\end{figure}

\begin{figure}
    \centering
    \includegraphics[width=1.0\linewidth]{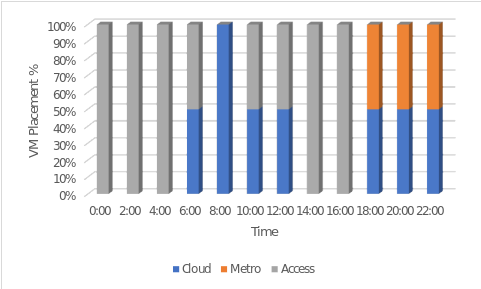}
    \caption{VM Placement in Cloud Fog under linear workload, 100 percent VM workload, and user download rate of 2 Mbps}
\end{figure}

\begin{figure}
    \centering
    \includegraphics[width=1.0\linewidth]{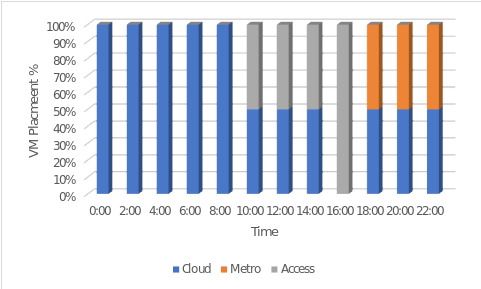}
    \caption{VM Placement in Cloud Green Fog under constant workload, 50 percent VM workload, and user download rate of 2 Mbps}
\end{figure}

\begin{figure}
    \centering
    \includegraphics[width=1.0\linewidth]{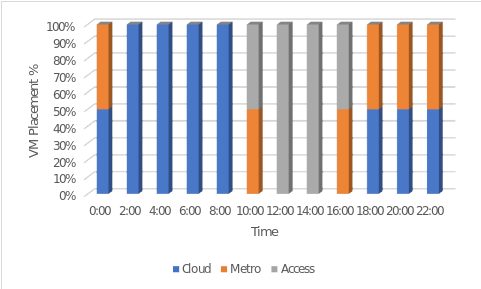}
    \caption{VM Placement in Cloud Green Fog under constant workload, 25 percent VM workload, and user download rate of 2 Mbps}
\end{figure}

\begin{figure}
    \centering
    \includegraphics[width=1.0\linewidth]{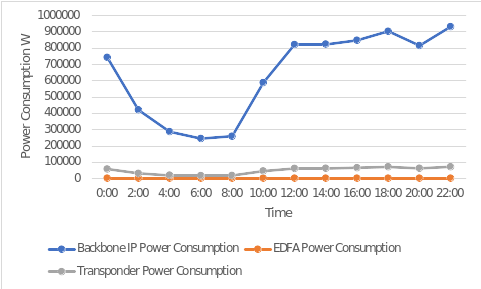}
    \caption{Overall Power consumption with only brown energy consideration}
\end{figure}

\begin{figure}
    \centering
    \includegraphics[width=1.0\linewidth]{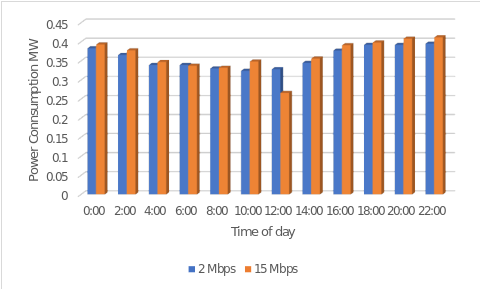}
    \caption{Overall Power consumption with solar energy consideration}
\end{figure}

\section{Conclusion}

There are clear advantages to the use of solar power in the fog layer, especially when considering virtual machine (VM) placement in a topology such as BT's 21st Century Network (21 CN). We have demonstrated that nodes powered by solar energy have been found to have more VM placement, therefore, providing lower latency rates to nearby users. This paper demonstrates one aspect of renewable energy solutions in network infrastructures by this notable improvement compared to other models in the field that solely depend on conventional 'brown' power sources.
\vspace*{50pt} 

\bibliography{solarvm}

\end{document}